\documentclass[11pt]{article}

\title{\bf Long Optimal or Small-Defect LRC Codes with Unbounded Minimum Distances}
\author{Hao Chen, Jian Weng and Weiqi Luo
  \thanks{Hao Chen, Jian Weng and Weiqi Luo are with the College of Information Science and Technology/Cyber Security, Jinan University, Guangzhou, Guangdong Province 510632, China, haochen@jnu.edu.cn, cryptjweng@gmail.com, lwq@jnu.edu.cn. The research of Hao Chen was supported by NSFC Grants 11531002. The research of Jian Weng was supported by NSFC Distinguished Young Scholar Grant 61825203. The research of Weiqi Luo was supported by NSFC Grant 61877029.}}

\begin{document}

\maketitle
\begin{abstract}
For a linear locally recoverable code (LRC) with length $n$, dimension $k$ and locality $r$ its minimum distance $d$ satisfies $d \leq  n-k+2-\lceil\frac{k}{r}\rceil$. A code attaining this bound is called optimal. Many families of optimal locally recoverable codes have been constructed by using different techniques in finite fields or algebraic curves. However in previous constructions of length $n>>q$ optimal LRC codes minimum distances are only few constants smaller than 9. No optimal LRC code over a general finite field ${\bf F}_q$ with the length $n\sim q^2$ and the minimum distance $d\geq 9$ has been constructed. In this paper we present a general construction of optimal LRC codes over arbitrary finite fields. Over any given finite field ${\bf F}_q$, for any given $r \in \{1,2,\ldots,q-1\}$ and given $d$ satisfying $3 \leq d \leq \min\{r+1,q+1-r\}$, we construct explicitly an optimal LRC code with length $n=q(r+1)$, locality $r$ and minimum distance $d$. We also gives an asymptotic bound for $q$-ary $r\leq q-1$-locality LRC codes better than the previously konwn bound. Many long $r$-locality LRC codes with small defect $s=n-k+2-\lceil \frac{k}{r} \rceil-d$ are also constructed.\\
\end{abstract}

\section{Introduction}

\subsection{Preliminaries}

Let ${\bf F}_q$ be finite field with $q$ elements where $q$ is a prime power. For a linear code ${\bf C}$ over ${\bf F}_q$ with length $n$, dimension $k$ and minimum distance $d$, we define the locality as follows. Given $a \in {\bf F}_q$, set ${\bf C}(i,a)=\{{\bf x} \in {\bf C}: x_i=a\}$, where $i \in \{1,\ldots,n\}$ is an arbitrary coordinate position. ${\bf C}_A(i,a)$ is the restriction of ${\bf C}(i,a)$ to the coordinate positions in $A \subset \{1,\ldots,n\}$. The linear code ${\bf C}$ is a locally recoverable code with locality $r$, if for each $i \in \{1,\ldots,n\}$, there exists a subset $A_i \in \{1,\ldots,n\}-\{i\}$ of cardinality at most $r$ such that ${\bf C}_{A_i}(i,a) \cap {\bf C}_{A_i}(i,a')=\emptyset$ for any given $a\neq a'$. It was proved a Singleton-like bound for LRC codes in \cite{GHSY2011} and \cite{PD2014} $$d \leq n-k+2-\lceil\frac{k}{r}\rceil,$$ where $\lceil x \rceil$ is the smallest integer greater than or equal to $x$. It is clear that $r \leq k$, this upper bound is just the Singleton bound $d \leq n-k+1$ for linear codes when $r=k$.  A linear code attaining this upper bound is called an optimal LRC code. In coding theory linear codes attaining the Singleton bound $d \leq n-k+2$ are called MDS (maximal distance separable). Then the optimal LRC code is a generalization of the MDS (Maximal Distance Separable) code.  We refer to \cite{GHSY2011,PD2014,TBF2016,YB2017} for the background in distributed storage.\\

The main conjecture of MDS codes claims that the length of an MDS code over ${\bf F}_q$ is at most $q+1$, except some trivial exceptional cases. Many optimal LRC codes with large code length $n>q$ have been constructed. Hence the main conjecture type upper bound on the lengths of optimal LRC codes does not hold directly. However it is still a challenging problem to ask the maximal possible length of an optimal LRC code over any given finite field ${\bf F}_q$. We refer to \cite{BHHMV2017, TB2014} for the discussion of the background. Considering the recent progress in \cite{BHHMV2017,Guruswami,LXC2018,XC,Jin} it is natural to ask that if there exists optimal LRC $q$-codes with length $n \sim q^2$ and unbounded locality and unbounded minimum distances. In this paper we give an affirmative answer.\\

On the other hand almost and near MDS codes have been studied in coding theory. Almost MDS codes are linear $[n,k,d]_q$ codes satisfying $n-k+1-d=1$. An almost MDS codes with its dual also almost MDS is called near MDS. We refer to \cite{DL,DDK,DL1,DeBoer,Gulliver,FW,D} for almost and near MDS codes. For near MDS codes it was conjectured that their maximal lengths are around $q+2\sqrt{q}$ (see \cite{D}). In \cite{FW} Corollary 8 it was proved that the length $n$ of an almost MDS $[n, k, n-k]_q$ code satisfies $n \leq d+2q$ if $k \leq 2$ and $d >q$. In this paper we call $s=n-k+2-\lceil \frac{k}{r} \rceil-d$ the defect of the locality-$r$ LRC code. Many long $r$-locality LRC codes with very large lengthes and small defects are also constructed, see Table 2.\\

The more general locally recoverable codes tolerating multiple erasures can be defined as follows. A linear code ${\bf C} \subset {\bf F}_q^n$ has $(r, \mu)$-locality if each coordinate position $i \in \{1,\ldots,n\}$ is contained in a subset $A \subset \{1,\ldots,n\}$ with cardinality $r+\mu-1$ such the restriction ${\bf C}_{A}$ of ${\bf C}$ to $A$ has minimum distance at least $\mu$. In the case $\mu=2$, it is just the LRC code with the locality $r$. The Singleton-like bound for a linear $[n,k,d]_q$ code with $(r,\mu)$-locality is $$d \leq n-k+1-(\lceil\frac{k}{r}\rceil-1)(\mu-1)$$  We refer the detail to \cite{PKLK2012,BTV2017}. A code attaining this bound is called an optimal LRC code with $(r,\mu)$-locality. Tamo-Barg good polynomial construction in \cite{TB2014} of $r$-LRC codes can be generalized to optimal LRC codes with $(r,\mu)$-locality. Some other optimal LRC codes with $(r,\mu)$-locality were constructed in \cite{PKLK2012,TBGC2016,CXHF2018,HHW}.\\

\subsection{Known LRC constructions}

We summarize previous constructions of optimal LRC codes in \cite{PKLK2012, WEH2014, GC2014,Guruswami,SRKV2013, TB2014, TBGC2016, BHHMV2017, LMC2018, LMX2018,CXHF2018,HHW} as follows.\\

\subsubsection{$n>q$}

1. {\bf Binary LRC codes over ${\bf F}_2$ with large lengths:} In \cite{WEH2014} an almost optimal binary LRC code with $n=15, k=10, r=6$ and $d=4<15-10+2-\lceil\frac{10}{6}\rceil=5$ was constructed. In \cite{GC2014} a family of optimal binary cyclic LRC codes satisfying $n=2^m-1$ for some positive integer $m$, $r+1|n$, $d=2$ was constructed. In \cite{HYUS} some upper bounds on the minimum distances of LRC and constructions of binary LRC were given. Many interesting construction of LRC with small localities over binary or small fields were given in \cite{ZY}.\\

2. {\bf Optimal LRC codes over ${\bf F}_4$ and ${\bf F}_5$ with large lengths:} In \cite{WEH2014} optimal LRC codes over ${\bf F}_4$ with $n=4i+4,k=3i+1,r=3, d=4$, $i \geq 1$ were constructed. In \cite{BHHMV2017} an optimal LRC code over ${\bf F}_4$ with $n=18, k=11, r=2, d=3$ was constructed. An optimal LRC code over ${\bf F}_5$ with $n=24, k=17, r=3, d=3$  and some other optimal LRC codes over ${\bf F}_7$ and ${\bf F}_{11}$ with length $48$ and $110$ were also constructed in \cite{BHHMV2017}. It was asked in \cite{BHHMV2017} if there exists a family of optimal LRC codes over ${\bf F}_q$ with length $n \sim q^2$, $d=3$ and all values of $r$. In \cite{LXC2018} distance $3$ and $4$ optimal LRC codes with arbitrary lengths were constructed via cyclic codes. The locality $r$ has to satisfy some number-theoretic property in the result of \cite{LXC2018}. \\

3. {\bf Optimal LRC codes over ${\bf F}_q$ with lengths up to $q+2\sqrt{q}$:} In \cite{LMX2018} by the using of elliptic curves and other algebraic-geometric techniques, optimal LRC codes over ${\bf F}_{p^a}$ with code length up to $p^a+2\sqrt{p^a}$ and locality $r \leq 5$ were constructed. In the case $q=2^a$, $a$ even, the locality of optimal LRC codes in \cite{LMX2018} can be $23$. To our knowledge this is the only known family of optimal LRC codes with larger distances over a general finite field with code lengths greater than field size. However the locality has to be smaller than or equal to $23$.\\

4. {\bf More general codes:} In July, 2018 Guruswami, Xing and Chen proved in \cite{Guruswami} an upper bound $n \leq O(dq^{3+\frac{4}{d-4}})$ on the length $n$ of an optimal LRC $[n,k,d]_q$ code over ${\bf F}_q$ satisfying $n \geq \Omega(dr^2)$. They also proved the existence of optimal LRC codes satisfying $n \geq \Omega_{d,r}(q^{1+\frac{1}{\lfloor \frac{d-3}{2}\rfloor}})$. In \cite{XC,Jin} some optimal LRC with $n$ very close to $q^2$ and small distances $d=5, 6, 7, 8$.\\

\subsubsection{$n<q$}

In \cite{PKLK2012} optimal LRC codes with $n=\lceil\frac{k}{r}\rceil(r+1)$ and $q>n$ was constructed. Optimal cyclic LRC codes over any given finite field ${\bf F}_q$ with $(r+1)|q-1$ and $n|q-1$ were constructed in \cite{TBGC2016}. In \cite{TB2014} optimal LRC codes over any given finite field ${\bf F}_q$ with $n$ slightly smaller than $q$ were constructed by the using of  good polynomials. This was extended in \cite{LMC2018} to give more such optimal LRC codes over any given finite field ${\bf F}_q$ with more possible values of the locality. In \cite{TBGC2016,CXHF2018,HHW, FF} optimal $(r,\mu)$-LRC codes with some special properties were constructed from cyclic codes. However few known optimal $(r,\mu)$-LRC codes over ${\bf F}_q$ have their lengths larger than $q$. {\em In all previous constructions minimum distances of long optimal LRC codes are bounded by some absolute constant.}\\

{\bf Main open problem.} In all above cases no optimal LRC code over ${\bf F}_q$ with length $n \sim q^2$ and and unbounded minimum distance $d \geq 9$ has been given.\\

In this paper we give an affirmative answer to this problem.\\

\subsubsection{Known asymptotic bound}.

In \cite{BTV2017} Proposition 6.3 the following asymptotic bound for limits $R=\lim \frac{k_i}{n_i}$ and $\delta=\lim \frac{d_i}{n_i}$ for sequences of LRC codes with locality $r$ and lengths $n_i \longrightarrow \infty$ was established by using Garcia-Stichtenoth curves. $$R \geq \frac{r}{r+1}(1-\delta-\frac{1}{\sqrt{q}+3}), r=\sqrt{q}-1,$$ and $$R \geq \frac{r}{r+1}(1-\delta-\frac{2\sqrt{q}}{q-1}), r=\sqrt{q}.$$  We will give new constructions of LRC codes with any given locality $r \leq q$ with better asymptotic bound in Section 4.\\

\subsection{Our contribution and an open problem}

In this paper we prove the following main result.\\

{\bf Main result.} {\em Over any given finite field ${\bf F}_q$, for a given positive integer $r \in \{1,2,\ldots, q-1\}$, a given positive integer $w$ satisfying $w \leq \min\{r-1,q-1-r\}$, and a positive integer $l \leq q$, an optimal LRC $[(r+1)l,rl-w,w+2]_q$ code with the locality $r$ can be constructed.}\\

We also extend our result to optimal $(r,\mu)$-LRC codes.\\

{\bf Corollary 1.1.} {\em 1) For any prime power $q \geq 7$, we construct explicit $[q^2-4q, q^2-5q-6, 8]_q$ optimal LRC code with locality $r=q-5$ over ${\bf F}_q$;\\
2) For any prime power $q\geq 7$, we construct explicit $[q^2-5q, q^2-6q-7, 9]_q$ optimal LRC code with locality $r=q-6$ over ${\bf F}_q$;\\
3) For any prime power $q\geq 13$, we construct explicit $[q^2-10q, q^2-11q-10, 12]_q$ optimal LRC code with locality $r=q-11$ over ${\bf F}_q$.}\\

Compared with known optimal LRC codes in subsection 1.2.1 our construction gives a lot of longer optimal LRC codes with unbounded minimum distances. This shows that the lengths of optimal LRC codes can be very close to $q^2$ even with unbounded minimum distances. This is quite different to the main conjecture type upper bound on MDS codes.\\

{\bf Open Problem.} From our construction and the result in \cite{Guruswami} it is natural to ask if there exist LRC codes with length $n \geq 2q^2$ and unbounded localities and unbounded distances.\\

We give a new asymptotic bound as follows. It is better than Proposition 6.3 in \cite{BTV2017} in some interval of $\gamma$.\\

{\bf Better asymptotic bound.} {\em Let $q$ be a square of a prime power. For any given locality $r \leq q-1$, and any given real number $\gamma \in (0,1)$, we have a sequence of $q$-ary LRC codes with locality $r$ satisfying $$\delta \geq (1-\gamma)(1-\frac{r-1}{q}),$$ and $$R \geq (\gamma-\frac{1}{\sqrt{q}})\frac{r}{q}$$.}\\

\section{Our construction}

\subsection{LRC codes}

Let ${\bf X}$ be a set and ${\bf F}_q$ be any given finite field. The function $g$ is a function ${\bf X} \longrightarrow {\bf F}_q$ such that there exist $l$ sets $A_1 \subset g^{-1}(y_1), \ldots, A_l \subset g^{-1}(y_l)$ of cardinality $|A_i|=r+1$, $i=1,\ldots,l$, where $y_1,\ldots y_l \in {\bf F}_q$ are $l$ distinct elements in ${\bf F}_q$ (then $l \leq q$). It is easy to construct the set ${\bf X}$ and the function $g$ satisfying the above property. For example, ${\bf X}=A \times {\bf F}_q$, where $A$ is a set of cardinality $r+1$ and $g$ is the projection to the second factor. For any given $${\bf a}=(a_{hs})_{0 \leq h \leq r-1,0 \leq s \leq t-1} \in {\bf F}_q^{k},$$ where $k=rt, t \leq l$, we consider the function $$F_{{\bf a}}(x, y)=\Sigma_{h=0}^{r-1}\Sigma_{s=0}^{t-1} a_{hs}g(x)^s y^h$$ on ${\bf X} \times {\bf F}_q$. Let ${\bf B} \subset {\bf F}_q$ be a subset with $r+1$ distinct elements $b_1,b_2,\ldots,b_{r+1}$ (then $r \leq q-1$) in ${\bf F}_q$. We denote $r+1$ elements of $A_i$ as $x_1^i,x_2^i,\ldots,x_{r+1}^i$, $i=1,\ldots,l$, then $g(x_j^i)=y_i$ for $j=1,\ldots,r+1$. The subset ${\bf A} \subset {\bf X} \times {\bf F}_q$ consists of the following $(r+1)l$ elements $(x_j^i,b_j)$ for $1\leq i \leq l$ and $1 \leq j \leq r+1$.\\

{\bf Proposition 2.1.} {\em We assume $t \leq l$. If $F_{{\bf a}}(x,y)$ is zero on all points of the set ${\bf A}$, then ${\bf a}=0$.}\\

{\bf Proof.} We consider $F_{{\bf a}}(x,y)$ on the subset consisting of $r+1$ elements $(x_1^i,b_1),\ldots,(x_{r+1}^i,b_{r+1})$. Then $$F_{{\bf a}}(x_j^i,b_j)=\Sigma_{h=0}^{r-1}(\Sigma_{s=0}^{t-1}a_{hs}g(x_j^i)^s)b_j^h.$$ Since $g(x_j^i)=y_i$, this is a constant for all $x_1^i,\ldots,x_{r+1}^i$, set $\Sigma_{s=0}^{t-1}a_{hs}g(x_j^i)^s=c_h$. Then the polynomial $\Sigma_{h=0}^{r-1} c_h x^h$ has $r+1$ roots $b_1,\ldots, b_{r+1}$. This implies that $\Sigma_{s=0}^{t-1} a_{hs} y_i^s=0$ for all possible $y_1,\ldots,y_l$. Since $t-1<l$, then the conclusion $a_{hs}=0$ follows directly.\\

\subsection{Recover procedure and minimum distance}

Set ${\bf U} \subset {\bf F}_q^{rt}$ be a linear subspace with dimension $u$, we consider the linear code ${\bf C(U)} \subset {\bf F}_q^n$, $n=(r+1)l$, defined by $${\bf C(U)}=\{(F_{{\bf a}}(x_j^i,b_j))_{1\leq j \leq r+1, 1\leq i \leq l}:{\bf a} \in {\bf U}\}.$$ Since $F_{{\bf a}}+F_{{\bf b}}=F_{{\bf a}+{\bf b}}$ and $F_{\lambda {\bf a}}=\lambda F_{{\bf a}}$ for $\lambda \in {\bf F}_q$ and ${\bf a}, {\bf b}$ in ${\bf F}_q^k$, this is a linear code with dimension $u$ from Proposition 2.1.\\

{\bf Definition 2.1.} {\em For any given ${\bf a} \in {\bf F}_q^k$, $k=rt$, $H_{{\bf a}}$ is number of common roots in the set {\bf B} of $t$ equations $\Sigma_{h=0}^{r-1} a_{hs}b^h=0$ for $s=0,1,\ldots,t-1$. That is, there exist $H_{{\bf a}}$ elements of the set ${\bf B}$, $b_{i_1},\ldots, b_{i_{H_{{\bf a}}}}$, such that $$\Sigma_{h=0}^{r-1} a_{hs}b_{i_f}^h=0$$ for $f=1,\ldots,H_{\bf a}$ and $s=0,1,\ldots,t-1$. We define $H({\bf U})=\max\{H_{\bf a}: {\bf a} \in {\bf U}\}$.}\\

{\bf Theorem 2.1.} {\em The locality of ${\bf C(U)}$ is at most $r$, the minimum distance of ${\bf C(U)}$ is at least $n-(r+1)(t-1)-H({\bf U})(l-t+1)$.}\\

{\bf Proof.} For a given coordinate position, say $(x_1^i, b_1)$, if $F_{{\bf a}^1}(x_1^i, b_1) \neq F_{{\bf a}^2}(x_1^i,b_1)$, then the evaluation vector of $F_{{\bf a}^1}$ and $F_{{\bf a}^2}$ at coordinate positions $(x_2^i,b_2),\ldots, (x_{r+1}^i, b_{r+1})$ can not be the same. Otherwise $$F_{{\bf a}^w}(x_j^i, b)=\Sigma_{h=0}^{r-1} (\Sigma_{s=0}^{t-1}a_{hs}^wg(x_j^i)^s)b^h=\Sigma_{h=0}^{r-1} (\Sigma_{s=0}^{t-1}a_{hs}^wy_i^s)b^h$$ are the same for $w=1,2$. Here we notice that $g(x_j^i)=y_i$ for all $j=1,2,\ldots,r+1$. If the evaluation vectors above are the same, since $(\Sigma_{s=0}^{t-1}a_{hs}^wy_i^s)$ are constants only depending on $i$ and ${\bf a}^w$, the two polynomials in $b$ of degree $r-1$ are the same at $r$ points $b_2,\ldots,b_{r+1}$. Then the two polynomials have to be the same, that is, $\Sigma_{s=0}^{t-1}a_{hs}^1y_i^s=\Sigma_{s=0}^{t-1}a_{hs}^2y_i^s$ for all $h=0,\ldots,r-1$. Then we have $F_{{\bf a}^1}(x_1^i,b_1)=F_{{\bf a}^2}(x_1^i,b_1)$. On the other hand if the evaluation at the $r$ points $(x_2^i,b_2),\ldots, (x_{r+1}^i, b_{r+1})$, $F_{{\bf a}}(x_j^i,b_j)=\Sigma_{h=0}^{r-1} (\Sigma_{s=0}^{t-1}a_{hs} g(x_j^i)^s)b_j^h$ are given, then the $r$ coefficients $\Sigma_{s=0}^{t-1}a_{hs} g(x_j^i)^s=\Sigma_{s=0}^{t-1}y_i^s$ can be solved from the Vandermonde matrix. Then the value $$F_{{\bf a}}(x_1^i,b_1)=\Sigma_{h=0}^{r-1} (\Sigma_{s=0}^{t-1}a_{hs} y_i^s)b_1^h$$ can be recovered. Here we notice that $g(x_1^i)=y_i$ from the definition of the function $g$. This is essentially the same as the recover procedure in page 4663 of \cite{TB2014}. Thus the locality is at most $r$.\\

For any given ${\bf a}=(a_{hs})_{0\leq h \leq r-1,0\leq s\leq t-1} \in {\bf U} \subset {\bf F}_q^{rt}$, we consider $$F(x_j^i,b_j)=\Sigma_{s=0}^{t-1} (\Sigma_{h=0}^{r-1}a_{hs}(b_j)^h)g(x_j^i)^s.$$ From Definition 2.1 the equation $\Sigma_{h=0}^{r-1}a_{hs}(b_j)^h=0$ for $s=0,\ldots,t-1$, is valid for $H_{{\bf a}} \leq r-1$ elements $b_{i_1},\ldots b_{i_{H_{{\bf a}}}}$ in the set ${\bf B}=\{b_1,\ldots,b_{r+1}\}$. Then the number of zeros of $F$ in the set ${\bf A}$ is at most $$H_{\bf a}l+(r+1-H_{\bf a})(t-1)=H_{\bf a}(l-t+1)+(r+1)(t-1).$$ Actually for each $b_{i_1,},\ldots,b_{i_{H_{{\bf a}}}}$, there are $l$ solutions. For any element $b_j$ in the set ${\bf B}-\{b_{i_1},\ldots,b_{i_{H_{{\bf a}}}})$, $$F(x_j^i,b_j)=\Sigma_{s=0}^{t-1} (\Sigma_{h=0}^{r-1}a_{hs}(b_j)^h)g(x_j^i)^s=\Sigma_{s=0}^{t-1} (\Sigma_{h=0}^{r-1}a_{hs}(b_j)^h)y_i^s$$ is not a zero polynomial, then it has at most $t-1$ possibilities of the value $y_i$ satisfying  $$\Sigma_{s=0}^{t-1} (\Sigma_{h=0}^{r-1}a_{hs}(b_j)^h)y_i^s=0.$$ For each such $y_i$, $g(x_j^i)=y_i$ has at most one solution $x_j^i$, since $j$ is fixed. The conclusion is proved.\\

\subsection{Optimal LRC Code construction}

\subsubsection{$d=2$}

This case is trivial. In the case ${\bf U}={\bf F}_q^{rt}$, it is clear $H({\bf U})=r-1$. Then $d=n-(r+1)t+2-((r-1)(l-t+1)-r+1)$. When $t<l$, $d=n-k+2-\lceil rt/r \rceil -((r-1)(l-t+1)-r+1)<n-k+2-\lceil rt/r \rceil$. When $t=l$, $d=n-(r+1)t+2$. That is we have a length $n=(r+1)l$, dimension $k=rl$, minimum distance $ d=2$ optimal LRC code over ${\bf F}_q$ with any given locality $r\leq q-1$.\\

\subsubsection{d=3}

{\bf Lemma 3.1.} {\em Let ${\bf F}_q$ be a finite field satisfying $q>\displaystyle{r+1 \choose r-w}$. For $\displaystyle{r+1 \choose r-w}$ linear subspaces $W_i\times \cdots \times \times W_i$ ($l$ copies) in ${\bf P}^{r-1}({\bf F}_q)\times \cdots \times{\bf P}^{r-1}({\bf F}_q)$ ($l$ copies), where $W_i$ is a dimension $w-1$ linear subspace in ${\bf P}^{r-1}({\bf F}_q)$, $i=1,\ldots,\displaystyle{r+1 \choose r-w}$, we can find a codimension $w$ linear subspace ${\bf U}$ in ${\bf P}^{r-1}({\bf F}_q)\times \cdots \times{\bf P}^{r-1}({\bf F}_q)$ ($l$ copies) such that the intersection of ${\bf U}$ with the union of these dimension $v$ linear subspaces is empty.}\\

{\bf Proof.} Suppose linear independent vectors have been chosen, in the final step, if $\displaystyle{r+1 \choose r-w} q^{r-1}<q^r$, then we can find the desired linear independent vector in ${\bf F}_q^r$. The conclusion is proved.\\

It should be noticed that when $q$ is small, the conclusion of Lemma 3.1 is not valid. For example if $r-1>q$, ${\bf P}^{r-1}({\bf F}_q)$ is covered by $\frac{r(r+1)}{2}$ hyper-planes defined $x_i=x_j$, where $i \neq j$.\\

In the above construction if ${\bf U}$ is the full space ${\bf F}_q^{rt}$, it is clear that $H({\bf U})$ can attain the maximal possibility $r-1$, that is, for some ${\bf a}=(a_{hs})$, there are $r-1$ elements $b$ in the set $\{b_1,\ldots,b_{r+1}\}$, such that $\Sigma_{h=0}^{r-1} a_{hs}b^h=0$ for all $s=0,1,\ldots,t-1$. Then in this case the minimum distance of the constructed optimal LRC code is $d=2$. Then we show that $d$ of the constructed optimal LRC code can be enhanced if $q>\displaystyle{r+1,\choose r+2-d}$ is satisfied.\\

For a equation $$\Sigma_{h=0}^{r-1}c_hb^h=0,$$ where $(c_0,c_1,\ldots,c_{r-1})$ are $r-1$ constant coefficients in ${\bf F}_q$ considered as a point in the projective space ${\bf P}^{r-1}({\bf F}_q)$, if $r-1$ roots are fixed, then $(c_0,c_1,\ldots,c_{r-1})$ is a fixed point in ${\bf P}^{r-1}({\bf F}_q)$. Then for $\frac{r(r+1)}{2}$ possibilities of $r-1$ roots in the set ${\bf B}=\{b_1,\ldots,b_{r+1}\}$, they corresponds to $\frac{r(r+1)}{2}$ points of coefficients in ${\bf P}^{r-1}({\bf F}_q)\times \cdots \times{\bf P}^{r-1}({\bf F}_q)$ ($l$ copies) satisfying that $\Sigma_{h=1}^{r-1}c_{hs}b_{i_j}^h=0$ for $s=0,1,\ldots,l-1$ and $j=1,\ldots,r-1$, where $(b_{i_1},\ldots,b_{i_{r-1}})$ is any fixed $r-1$ elements in the set ${\bf B}$. From Lemma 3.1 if $q> \displaystyle{r+1 \choose r-1}$, there exists a codimension $1$ linear subspace ${\bf U}$ of ${\bf P}^{r-1}({\bf F}_q)\times \cdots \times{\bf P}^{r-1}({\bf F}_q)$ ($l$ copies) such that these $\frac{r(r+1)}{2}$ coefficient points are not in ${\bf U}$. That is for ${\bf a} \in {\bf U} \subset {\bf F}_q^{rl}$, at least for one of $s \in \{0,1,\ldots,t-1\}$, $$\Sigma_{h=0}^{r-1} a_{hs}b^h=0$$ can not have $r-1$ roots in the set ${\bf B}$. Hence $H({\bf U})$ in Theorem 2.1 can not be its maximal possibility $r-1$, we have $H({\bf U}) \leq r-2$. From ${\bf U}$ we have a linear code with length $n=(r+1)l$, dimension $k=rl-1$ and distance $d \geq n-(r-2)l-(r+1-r+2)(l-1)=3$. This code has locality $r$ and satisfies the Singleton-like bound. It is an optimal LRC code with $n=(r+1)l,k=rl-1, d=3$ and locality $r$.\\

\subsubsection{General case}

In general we consider the case that the equation $$\Sigma_{h=0}^{r-1}c_hb^h=0,$$ where $(c_0,c_1,\ldots,c_{r-1})$ are $r-1$ constant coefficients in ${\bf F}_q$ considered as a point in the projective space ${\bf P}^{r-1}({\bf F}_q)$, has at least $r-w$ roots in the set ${\bf B}$, where $w$ is a fixed positive integer in the set $\{1,2,\ldots,r-1\}$. For example, suppose $b_1,\ldots,b_{r-w}$ are such $r-w$ roots. Then the equation is of the form $$C(x-b_1)\cdots (x-b_{r-w})(x^{w-1}+c'_{w-2}x^{w-2}+\cdots+c'_1x+c'_0)$$, where $C,c'_{w-1},\ldots,c'_1,c'_0$ are $w+1$ variables. Then the coefficient points $(c_0,\ldots,c_{r-1})$ corresponds to a linear subspace in ${\bf P}^{r-1}({\bf F}_q)$ of dimension $w-1$. We have $\displaystyle{r+1 \choose r-w}$ such linear subspaces in ${\bf P}^{r-1}({\bf F}_q)\times \cdots \times{\bf P}^{r-1}({\bf F}_q)$ ($l$ copies). Hence from Lemma 3.1 if $q> \displaystyle{r+1 \choose r-w}$, a linear subspace in ${\bf P}^{r-1}({\bf F}_q)\times \cdots \times{\bf P}^{r-1}({\bf F}_q)$ ($l$ copies) with codimension $w$ can be found which has no intersection with these $\displaystyle{r+1 \choose r-w}$ products of such linear subspaces. That is, we can find a linear subspace ${\bf U}_w$ of ${\bf P}^{r-1}({\bf F}_q)\times \cdots \times{\bf P}^{r-1}({\bf F}_q)$ ($l$ copies) of codimension $w$ such that for ${\bf a} \in {\bf U}_w \subset {\bf F}_q^{rl}$, at least for one of $s \in \{0,1,\ldots,t-1\}$, $$\Sigma_{h=0}^{r-1} a_{hs}b^h=0$$ can not have $r-w$ roots in the set ${\bf B}$. Then $H({\bf U}_w) \leq r-w-1$. The Singleton-like upper bound is $d \leq (r+1)l-(rl-w)+2-\lceil\frac{rl-w}{r}\rceil=w+2$. From Theorem 2.1, the lower bound is $d \geq (r+1)l-(r+1)(l-1)-(r-w-1)=w+2$. Hence the locality is $r$ and $d=w+2$.\\

Therefore we have proved that the minimum distance of the constructed optimal LRC code can be enhanced when the field size is large. In the following part we prove that actually the same conclusion can be proved when the field size  satisfies $q\geq r+d-1$.\\

\subsubsection{Construction based on the Vandermonde matrix}

The above construction depends on Lemma 3.1 with a "counting points" argument. However the linear subspace of ${\bf P}^{r-1}({\bf F}_q)$ (of the coefficients $(c_0,\ldots, c_{r-1})$ ) defined by the condition that "there are $r-w$ roots $b_{i_1},\ldots,b_{i_{r-w}}$ in the set ${\bf B}$ " is defined by a $(r-w) \times r$ partial Vandermonde matrix as follows.\\
$$
\left(
\begin{array}{cccccc}\\
1&b_{i_1}&\cdots&b_{i_1}^{r-1}\\
1&b_{i_2}&\cdots&b_{i_2}^{r-1}\\
\cdots&\cdots&\cdots&\cdots\\
1&b_{i_{r-w}}&\cdots&b_{i_{r-w}}^{r-1}\\
\end{array}
\right)
$$
Hence if $r+1+w \leq q$ is satisfied, we can pick up $w$ distinct elements $e_1,\ldots,e_w$ in ${\bf F}_q-{\bf B}$. Then the codimension $w$ linear subspace ${\bf U}$ in ${\bf P}^{r-1}({\bf F}_q)$ defined by the following partial $w \times r$ Vandermonde matrix satisfying the requirement in Lemma 3.1. Actually a $r \times r$ Vandermonde matrix from $r$ {\bf distinct} elements is of rank $r$.\\
$$
\left(
\begin{array}{cccccc}\\
1&e_1&\cdots&e_1^{r-1}\\
1&e_2&\cdots&e_2^{r-1}\\
\cdots&\cdots&\cdots&\cdots\\
1&e_w&\cdots&e_w^{r-1}\\
\end{array}
\right)
$$
The two conditions about the coefficient vector that\\
1)there are $r-w$ roots in the set ${\bf B}$ and\\
2)in the subspace ${\bf U}$,\\
correspond to a rank $r$ Vandermonde $r\times r$ matrix. If the linear subspace ${\bf U}$ has non-empty intersection with one of the $\displaystyle{r+1 \choose r-w}$ linear subspaces in Lemma 3.1, the point in the intersection has to be a zero vector. This is a contradiction. Then we have the following result.\\

{\bf Theorem 2.2.} {\em For any given finite field ${\bf F}_q$, a positive integer $r \in \{1,2,\ldots,q-1\}$, a positive integer $1 \leq w \leq q-1-r$, and a positive integer $l \leq q$, an optimal LRC $[(r+1)l,rl-w,w+2]_q$ code with the locality $r$ can be constructed.}\\

In the following Table 1 we give many long optimal LRC codes with $n$ very close to $q^2$.\\

\begin{center}
{\bf Table 1} Explicit optimal LRC codes with $n \sim q^2$\\
\bigskip
\begin{tabular}{||c|c||c|c||}\hline
locality& length &dimension &distance \\ \hline
$q-3$&$q^2-2q$&$q^2-3q-4$&6\\ \hline
$q-4$&$q^2-3q$&$q^2-4q-5$&7\\ \hline
$q-5$&$q^2-4q$&$q^2-5q-6$&8\\ \hline
$q-6$&$q^2-5q$&$q^2-6q-7$&9\\ \hline
$q-7$&$q^2-6q$&$q^2-7q-8$&10\\ \hline
$q-8$&$q^2-7q$&$q^2-8q-9$&11\\ \hline
$q-9$&$q^2-8q$&$q^2-9q-10$&12\\ \hline
$q-10$&$q^2-9q$&$q^2-10q-11$&13\\ \hline
\end{tabular}
\end{center}

\section{Long optimal $(r,\mu)$-LRC codes}

Let $\mu \geq 2$ be a positive integer. We construct a function $g$ as in section 2 and  pick up $l$ subsets $A_i \subset g^{-1}(y_i)$, where $y_1,\ldots,y_l$ are distinct $l$ elements in ${\bf F}_q$, $|A_i|=r+\mu-1$, $i=1,2,\ldots,l$. For any given $${\bf a}=(a_{hs})_{0 \leq h \leq r-1,0 \leq s \leq t-1} \in {\bf F}_q^{k},$$ where $k=rt, t \leq l$, we consider the function $$F_{{\bf a}}(x, y)=\Sigma_{h=0}^{r-1}\Sigma_{s=0}^{t-1} a_{hs}g(x)^s y^h$$ on ${\bf X} \times {\bf F}_q$. Let ${\bf B} \subset {\bf F}_q$ be a subset with $r+\mu-1$ distinct elements $b_1,b_2,\ldots,b_{r+\mu-1}$, then $r+\mu \leq q$. We denote $r+\mu-1$ elements of $A_i$ as $x_1^i,x_2^i,\ldots,x_{r+\mu-1}^i$, $i=1,\ldots,l$. The subset ${\bf A} \subset {\bf X} \times {\bf F}_q$ consists of the following $(r+\mu-1)l$ elements $(x_j^i,b_j)$ for $1\leq i \leq l$ and $1 \leq j \leq r+\mu-1$. Set ${\bf U} \subset {\bf F}_q^{rt}$ be a linear subspace with dimension $u$, we consider the linear code ${\bf C(U)} \subset {\bf F}_q^n$, where $n=(r+\mu-1)l$, defined by $${\bf C(U)}=\{(F_{{\bf a}}(x_j^i,b_j):i=1,\ldots,l,j=1,\ldots, r+\mu-1): {\bf a} \in {\bf U}\}.$$ This is a linear code with dimension $u$. We have the following result.\\

{\bf Theorem 3.1.} {\em ${\bf C(U)}$ is a $(r,\mu)$-LRC code, the minimum distance of ${\bf C(U)}$ is at least $n-(r+\mu-1)t+\mu-(H({\bf U})(l-t+1)-r+1)$.}\\

{\bf Proof.} We consider the restriction of ${\bf C(U)}$ to the subset $${\bf B}_i=\{(x_1^i,b_1),\ldots,(x_{r+\mu-1}^i,b_{r+\mu-1})\}$$ Then the conclusion follows from a similar argument as the proof of Theorem.2.1.\\

When $t=l,{\bf U}={\bf F}_q^{rl}$, then $H({\bf U})=r-1$. Hence we get a $(r,\mu)$-LRC code attaining the Singleton-like bound, with length $n=(r+\mu-1)l, k=rl,d=\mu$. Hence for any given finite field ${\bf F}_q$, a positive integer $l \leq q$, a locality $(r,\mu)$ where $r$ is any value in $\{1,2,\ldots,q-1\}$ and $\mu$ is an arbitrary positive integer satisfying $2\leq \mu \leq q+1-r$, an optimal $(r,\mu)$-LRC code with length $n=(r+\mu-1)l$, dimension $k=rl$ and minimum distance $d=\mu$ can be constructed.\\

For optimal $(r,\mu)$-LRC codes we have the following result by a similar construction as in the proof of Theorem 2.2.\\

{\bf Corollary 3.1.} {\em For any given finite field ${\bf F}_q$, any given $(r,\mu)$ satisfying $r+\mu \leq q+1$, a positive integer $w$ satisfying $1\leq w \leq q+1-r-\mu$, a positive integer $l\leq q$, an optimal $(r,\mu)$-LRC code over ${\bf F}_q$ with length $(r+\mu-1)l$, dimension $rl-w$, and distance $w+\mu$ can be explicitly constructed.}\\

Hence many optimal $(r, \mu)$-LRC codes with lengths $n \sim q^2$ are constructed.\\

\section{Better asymptotic bound}

Let ${\bf X}$ be a smooth projective absolutely irreducible curve of genus $g$ defined over ${\bf F}_q$, ${\bf P}=\{P_1,\ldots,P_n\}$ be a set of $n$ ${\bf F}_q$-rational points, ${\bf G}$ be a ${\bf F}_q$-rational divisor with its degree satisfying $\deg({\bf G}) <n$. Let $f_0,\ldots,f_{t-1}$ be the base of $L({\bf G})=\{f: div (f)+{\bf G} \geq 0\}$, where $t=\deg({\bf G})-g+1$. For any given value $r \leq q$, we consider functions $$F({\bf a},x,b)=\Sigma_{h=0}^{r-1}\Sigma_{s=0}^{t-1}a_{hs}f_s(x)b^h,$$ defined on ${\bf X} \times {\bf F}_q$ where $(x,b) \in {\bf X} \times {\bf F}_q$ and ${\bf a}=(a_{hs})_{0 \leq h \leq r-1, 0 \leq s \leq t-1} \in {\bf F}_q^{rt}$. For a given linear subspace ${\bf U}$ of dimension $u$ in ${\bf F}_q^{rt}$, we define a linear code $C({\bf U})=\{(F({\bf a},x,b)): (x,b) \in {\bf P}\times{\bf F}_q, {\bf a} \in {\bf U}\}$. Since $r-1\leq q$ and $\deg({\bf G}) <n$, the evaluation mapping on ${\bf P} \times {\bf F}_q$ is injective, this is a dimension $u$ linear code with length $nq$ and dimension $k=(\deg({\bf G})-g+1)r$.\\

{\bf Theorem 4.1.} {\em The minimum distance of ${\bf C(U)}$ is at least $(n-\deg({\bf G}))(q-r+1)$. Suppose $r \leq q-1$, this code ${\bf C(U)}$ has locality $r$.}\\

{\bf Proof}. For each $$F({\bf a},x,b)=\Sigma_{h=0}^{r-1}\Sigma_{s=0}^{t-1}a_{hs}f_s(x)b^h=\Sigma_{h=0}^{r-1}(\Sigma_{s=0}^{t-1}a_{hs}f_s(x))b^h$$ for $x \in {\bf P}$ and $b \in {\bf F}_q$, there are at most $\deg({\bf G})$ points in ${\bf P}$ such that the coefficients $\Sigma_{s=0}^{t-1}a_{hs}f_s(x)$ are all zero. For the remaining $n-\deg({\bf G})$ points in ${\bf P}$, the degree $r-1$ non-zero polynomial $F({\bf a},x,b)$ of $b$ has at most $r-1$ roots in ${\bf F}_q$. Then $F({\bf a},x,b)$ as a whole has at most $\deg({\bf G})q+(n-\deg({\bf G}))(r-1)$ zero points in ${\bf P} \times {\bf F}_q$. The minimum distance is at least $$nq-\deg({\bf G})q-(n-\deg({\bf G}))(r-1)=(n-\deg({\bf G}))(q-r+1).$$

For any given point, f.g., $(P_1,b_1) \in {\bf P} \times {\bf F}_q$, we pick up $r$ distinct points $(P_1,b_1),\ldots,(P_1,b_r)$ in ${\bf P}\times {|bf F}_q$, where $b,b_1,\ldots,b_r$ are $r+1$ distinct elements of ${\bf F}_q$ since $r \leq q-1$. If two functions $F({\bf a}^1,x,b)$ and $F({\bf a}^2,x,b)$ are the same at the $r$ points $(P_1,b_1),\ldots,(P_1,b_r)$, then two degree $r-1$ polynomials $$F({\bf a}^1,P_1,b)=\Sigma_{h=0}^{r-1}(\Sigma_{s=0}^{t-1}a_{hs}^1f_s(P_1))b^h$$ and $$F({\bf a}^2,P_1,b)=\Sigma_{h=0}^{r-1}(\Sigma_{s=0}^{t-1}a_{hs}^2f_s(P_1))b^h$$ of $b$ are the same at $r$ elements $b_1,\ldots,b_r$ of ${\bf F}_q$. Hence they have the same coefficients $$\Sigma_{s=0}^{t-1}a_{hs}^1f_s(P_1)=\Sigma_{s=0}^{t-1}a_{hs}^2f_s(P_1).$$ Therefore $$F({\bf a}^1,P_1,b)=F({\bf a}^2,P_1,b).$$  The recover procedure is the same as the proof of Theorem 2.1. The conclusion is proved.\\

We take the family of Garcia-Stichtenoth curves ${\bf X}_m$ of genus $g_m \longrightarrow \infty$ over ${\bf F}_q$ where $q$ is a square of a prime power, $$\lim\frac{g_m}{N_m}=\sqrt{q}.$$ Here $N_m$ is the number of ${\bf F}_q$ rational points of ${\bf X}_m$. In Theorem 4.1 set ${\bf P}_m$ as the set of all these ${\bf F}_q$ rational points. Then we have a family of $r$-locality LRC codes for any given $r \leq q-1$ with length $n_m=N_mq$, dimension $k_m=(\deg({\bf G}_m)-g_m+1)r$ and minimum distance $d_m \geq (N_m-\deg({\bf G}_m))(q-r+1)$. We take a sequence of ${\bf F}_q$ rational divisors ${\bf G}_m$ such that $\lim \frac{\deg({\bf G}_m)}{N_m}=\gamma$. The conclusion of the following Corollary 4.1 follows directly.\\

{\bf Corollary 4.1} {\em Let $q$ be a square of a prime power. For any given fixed locality $r \leq q-1$, and any given real number $\gamma \in (0,1)$, we have a sequence of $q$-ary LRC codes with locality $r$ satisfying $$\delta \geq (1-\gamma)(1-\frac{r-1}{q}),$$ and $$R \geq (\gamma-\frac{1}{\sqrt{q}})\frac{r}{q}.$$}\\

\section{Long LRC codes with small defects}

In the construction if we replace ${\bf F}_q$ by a subset ${\bf S} \subset {\bf F}_q$ of $r+1$ elements the following result follows directly.\\

{\bf Theorem 5.1.} {\em If $r \leq q-1$ is fixed, we construct an explicit length $n(r+1)$, dimension $(\deg({\bf G})-g+1)r$, minimum distance $d \geq 2(n-\deg({\bf G}))$ $r$-locality LRC code. The defect is $s=(n-\deg({\bf G}))(r-1)+(g-1)(r+1)+2$.}\\

{\bf Corollary 5.1.} {\em From a elliptic curve with $N$ rational points over ${\bf F}_q$ and a fixed $r \leq q-1$ and an integer $1 \leq t \leq N$, we construct $r$-locality LRC codes with length $(r+1)N$, dimension $tr$, minimum distance $2(N-t)$. The defect $s=N-t+2$.}\\

It follows Theorem 5.1 we have the following asymptotic bound for $r$-locality LRC codes with reasonable small defect limit $S =\lim\frac{s}{n(r+1)}$.\\

{\bf Corollary 5.2} {\em Let $q$ be a square of a prime power. For any given fixed locality $r \leq q$, and any given real number $\gamma \in (0,1)$, we have a sequence of $q$-ary LRC codes with locality $r$ satisfying $$\delta \geq \frac{2(1-\gamma)}{r+1},$$ $$R \geq (\gamma-\frac{1}{\sqrt{q}})\frac{r}{r+1}$$ and $$S \leq (1-\gamma)\frac{r-1}{r+1}+\frac{1}{\sqrt{q}}.$$}\\

In the following Table 2 $r$-locality LRC codes over small fields with large lengths and small defects are given. Algebraic curves with many rational points are from \cite{Geer}.\\

\begin{center}
{\bf Table 2} Explicit $r$-locality LRC codes over ${\bf F}_q$\\
\bigskip
\begin{tabular}{||c|c|c||c|c|c||}\hline
q&locality& length &dimension &distance &defect\\ \hline
3&2&21&8&6&5\\ \hline
3&2&24&8&6&8\\ \hline
3&2&30&8&8&12\\ \hline
3&2&36&10&8&15\\ \hline
4&2&27&8&10&7\\ \hline
4&2&30&8&10&16\\ \hline
8&3&56&36&4&6\\ \hline
8&3&96&60&4&14\\ \hline
8&4&70&48&4&6\\ \hline
8&4&120&80&4&16\\ \hline
9&3&64&42&4&6\\ \hline
9&3&64&36&8&10\\ \hline
9&3&112&69&6&16\\ \hline
16&3&100&60&10&7\\ \hline
27&3&152&90&16&10\\ \hline
32&3&176&120&8&6\\ \hline
64&3&324&232&8&6\\ \hline
81&3&400&270&20&12\\ \hline
128&3&600&420&20&12\\ \hline

\end{tabular}
\end{center}

\section{Conclusion}

In this paper for any given finite field ${\bf F}_q$, any given $r \in \{1,2,\ldots,q-1\}$ and given $d$ satisfying $3 \leq d \leq \min\{r+1,q+1-r\}$, we give an optimal LRC code with length $n=q(r+1)$, locality $r$ and minimum distance $d$. This is the only known family of optimal LRC codes with $n \sim q^2$, unbounded localities and unbounded distances. We speculate there exist optimal LRC codes with $n \geq 2q^2$, unbounded localities and unbounded distances. By the using of Garcia-Stichtenoth curves a better asymptotic bound for $r$-locality LRC codes is presented. We also construct many long $r$-locality LRC codes with small defects.\\

\end{document}